# Introducing ASTROMOVES

*Jarita Holbrook*[1,2]

1 University of Edinburgh, United Kingdom

2 University of the Western Cape, South Africa

Jc.holbrook@ed.ac.uk

**Abstract:** The ASTROMOVES project studies the career moves and the career decision-making of astrophysicists. The astrophysicists participating have to have made at least two career moves after receiving their doctorates, which is usually between 4 and 8 years post PhD. ASTROMOVES is funded via the European Union and thus each participant must have worked or lived in Europe. Gender, ethnicity, nationality, marital status, and if they have children are some of the many factors for analysis. Other studies of the careers of astronomers and astrophysicists have taken survey approaches (Fohlmeister & Helling, 2012, 2014; Ivie et al., 2013; Ivie & White, 2015) laying a foundation upon which ASTROMOVES builds. For ASTROMOVES qualitative interviews are combined with publicly available information for the project, rather than surveys. Valuable information about career options and the decisions about where not to apply will be gathered for the first time. Those few studies that have used qualitative interviews often include both physicists and astrophysicists, nonetheless they have revealed issues that are important to ASTROMOVES such as the role of activism and the nuances of having children related to the long work hours culture (Ong, 2001; Rolin & Vainio, 2011). The global COVID-19 pandemic has slowed down the project; however, at the time of this writing 20 interviews have been completed. These interviews support previous research findings on how having a family plays an important role in career decision making, as well as the importance of mobility in building a career in astrophysics. Unexpected preliminary results include imposter syndrome, unemployment, stalking and coping with the global pandemic (Holbrook, 2021). Cultural Astronomy spans all aspects of the relationship between humans and the sky as well as all times ancient to the present; and thus, studying astronomers & astrophysicists who have a professional relationship to the sky is part of cultural astronomy, too.

**Introduction: Astrophysicists on the Move**

As with many academic disciplines, for astrophysicists to secure an academic job requires a doctorate, experience doing independent research in the form of a postdoctoral position, then joining an academic faculty as a temporary member until being promoted to a permanent faculty position, and finally being promoted to the most senior rank. The faculty job market for astrophysicists is very small approximated at 180 permanent-track positions advertised per year (Jacobson-Galan, 2021). New doctorates in astrophysicists in very rare cases go straight from their PhD to a permanent-track faculty position, but are much more likely to do one or two postdocs before obtaining a permanent-track position. Postdoctoral positions very in length with the shortest being a year and the longest being five years, most fall somewhere in between at around three years. Just as postdocs very with appointment period, postdoctoral positions very in prestige and other benefits. For example, named fellowships are considered more prestigious and often come with higher salaries and generous travel and research funds. Having two postdocs means two changes of position and oftentimes that means relocating twice. More postdocs means more position changes and potentially more movement. How do astrophysicists navigate this tight job market? Do they and are they willing to move abroad as part of their career navigation?

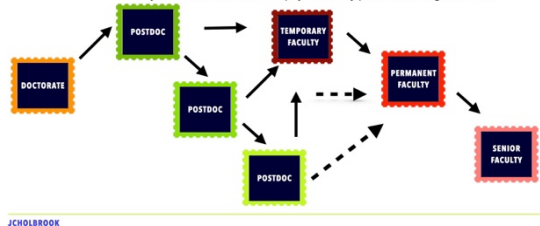

**Figure 1: Slide capturing the number of postdocs and positions that astrophysicists pass through in their efforts to obtain a permanent position. Created by JC Holbrook.**

The demographics of astrophysics are heavily weighted towards men of European descent all over the world. For example, in the United Kingdom's Royal Astronomical Society survey the survey respondents were 90% white and 66% male (Sean McWhinnie, 2017). In terms of the International Astronomical Union, 79% are male, however no race/ethnicity information is included (IAU, 2021). 'Intersectionality' is an idea/theory/theme/framework that has emerged from the works of Kimberley Crenshaw (Crenshaw, 1989, 1991). Crenshaw showed that people that embodied more than one marginalized identity were often overlooked/excluded/uncounted/discounted. For example, in the often-cited case against General Motors car company in the USA, white women were hired into office support positions, while Black men were hired into assembly line positions; however, Black women were excluded from both. The combination of Black and women led to an overlooked exclusion, though the legal case was lost, reflecting on the case lead to Crenshaw's development of Intersectionality. Astrophysicists often state that they seek to diversify the demographics of their community, but they have been not as successful as they expected in terms of seeing very little change over a long period of time; for example, the first women in astronomy conference was held in 1992 and the demographic statistics has changed little since then (C. Megan Urry, 1993; Hughes, 2014).

ASTROMOVES was designed to study intersectional identities and how astrophysicists navigate their careers in this difficult job market. The question is what are the experiences of astrophysicists embodying intersectional identities as they navigate their careers and how these compare to the majority group, men and men of European descent (white).

Previous studies of astrophysicists and their careers have focused on collecting data via surveys. These surveys lay a foundation for ASTROMOVES and helped shape the interview areas. Starting with surveys of women, in 2011 Fohlmeister & Helling (Fohlmeister & Helling, 2012) surveyed 61 women astrophysicists living in Germany. Only 8 held permanent positions. Important for ASTROMOVES, working abroad was listed as factor for advancing a career in astronomy. In 2014, they added a survey of 71 men working in Germany, 169 women working in the United Kingdom and 108 men working in the United Kingdom (Fohlmeister & Helling, 2014). Adding the survey results for men, changed the order of the factors important for navigating an astrophysics career, for example, 'vicinity of family' was #2 for German women, but this is #4 overall for the UK & German men and women. Ivie & White (Ivie & White, 2015) did a survey comparing the perception of research culture across 9 countries (Argentina, Canada, China, France, Germany, Italy, Japan, Spain and the United States of America); both men and women were surveyed. In terms of having children, women in all the countries surveyed were more likely to say that their career progression slowed once they had children. Whereas, men more likely to say their career did not change after having children. This is significant for ASTROMOVES as people that are married and have children were included in the study.

**Methodology**

The primary data source for ASTROMOVES are the astrophysicists, thus semi-structured interviewing was used. However, before conducting interviews decisions had to be made about the interviewees. To have interviews of reasonable length for analysis, each person needed to have changed position at least two times since they received their doctorate. Given the intersectional lens of the project, identifying people that had intersectional identities were targeted. Prior to starting the project, emails were sent to a selection of scientists that identified themselves as gender diverse on the Astronomy Outlist (Mao & Blaes, 1998) which continues to be updated. Twenty scientists were emailed and fourteen of these agreed that if the project moved forward, they were happy to be interviewed for the project. The heterosexual scientists were identified using a few techniques: convenance, targeted and snowball sampling. Needless to say, the original method of identifying scientists was changed due to the global COVID-19 pandemic. Instead, almost all of the interviews were conducted online using Zoom™. Zoom has to be mentioned because there was a benefit to using Zoom, it transcribed all the recorded interviews saving hours of labour. All interviews were conducted in English. It was proposed that 50 interviews would be completed for the project, however, the pandemic delayed the project start so it waits to be seen if that many interviews will be completed before the project ends.

The scientists chose among three different levels of consent from anonymous to pubic. The public option was for those scientists interested in being part of a proposed documentary film on the project. For articles and presentations each person was assigned or they chose a Hawaiian pseudonym. The pseudonym is how specific quotes are identified regardless of level of consent.

Interview transcripts were reformatted using their pseudonyms in preparation for qualitative analysis using NVIVO™. NVIVO was used solely because it is the qualitative analysis software for which the University of Edinburgh holds a site licence. The interviews were coded exploring multiple themes many of which emerged from the interviews (grounded theory).

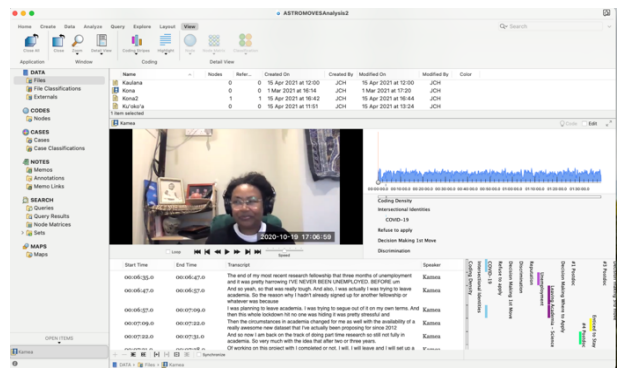

**Figure 2: Screen shot showing the NVIVO workspace on a Mac computer. Interview video and sound files are in the middle of the workspace. Transcript is below with coding stripes on the lower right.**

The interviews were contextualized by examining their associated CVs, their presence or absence on the astrophysics job rumour mill (Anonymous, 2020), and the American Astronomical Society's job register (American Astronomical Society, 2021). Most CVs were publicly available but several are private documents that were provided upon my request or my teams' request. Other textual data sources include email exchanges and notes (fieldnotes) written during the interviews.

**Research Results**

Thirty-five interviews have been completed. Of these the majority of heterosexual males (see Table 1), the scientists often combined gender, sex and sexuality in this way, often saying 'straight' rather than 'heterosexual'.

**Table 1: Gender and Sex. The category LGBTQIA+ includes everybody in terms of sex: males, females and transgender.**

| | | |
|---|---|---|
| Heterosexual Males | 15 | 43% |
| Heterosexual Female | 12 | 34% |
| LGBTQIA+ | 8 | 23% |
| Total | 35 | 100% |

The career age of a scientist was calculated in whole years from the date of obtaining their PhD to the present. Career age is more useful for analysis than actual age given the focus on career moves. Career age takes into account that scientists may have become students later in life, but does not take into account career breaks associated with childcare, eldercare or illness.

**Table 2: Career Age. Career age is calculated in whole years from the year of PhD to the present.**

| Career Age | Heterosexual Females | Heterosexual Males | LGBTQIA members |
|---|---|---|---|
| >20 | 3 | 5 | 1 |
| 10 to 19 years | 4 | 7 | 3 |
| <10 | 5 | 3 | 4 |
| Total | 12 | 15 | 8 |

The average career age per group is as follows with the error being the standard deviation. Heterosexual Females have an average career age of 13 ± 8 years, Heterosexual Males have an average of 17 ± 11 years, and LGBTQIA Members have an average career age of 12 ± 7 years. Thus, for this interview population of 35 scientists the Heterosexual Males are older than the other two groups.

Moving forward with analysis, these two aspects career age and gender sex are categories of analysis and comparison, for example when they achieve a permanent position or when family becomes part of their career decision-making. Such results will be presented in the future.

**Table 3: The importance of Science in the first career move.**

| First Position | Number of People |
|---|---|
| Mentioned that the Science was a factor | 28 |
| Other factors were more important such as the only postdoc offer, a place where they spoke the language, 2-body, near family… | 7 |

> My partner was very unhappy living in [European Country]. It was a very productive time for me in terms of getting to go to conferences and producing papers. I feel like my profile in my subfield grew at least in Europe… it's hard to stay connected to the American side of things, while living in Europe. But my partner was just very unhappy in Europe. So, then at that point, I had an opportunity to do a well-funded prestigious postdoc in__________ or I could do a smaller appointment at _____________, but it would be back in [their home city]. Because of things going on in my personal life, I decided 'Hey, let's go back to [our home city]!" - Palolo

Palolo is one of the scientists interviewed. Palolo was already married before moving to their first postdoc position in Europe, at that time their spouse was open for the adventure of moving to Europe. As the quote says, living in Europe was difficult for Palolo's partner. For their second postdoc, the happiness of Palolo's partner played a much greater role than for the first postdoc in their decision-making. Palolo implies that the choice made was not the best in terms of advancing their science or advancing their career, but instead was a return to familiar territory. What isn't said but is implied is that it was a return to established social networks of family and friends, as well.

Though still being analysed, similar to the findings previous surveys, the first postdoc was chosen because of the science (see **Error! Reference source not found.**) i.e. interesting science or the next logical step of their research or access to a useful dataset/telescope (Fohlmeister & Helling, 2012, 2014); however, other factors begin to be given greater weight as their career age increases. This time dimension hasn't been captured adequately in previous surveys.

Mapping career moves is not so much for analysis as another way of presenting complex information. Romeel Davé has given permission for their career map to be shown (see Figure 3). The USA and UK are well known places for astrophysics, however South Africa, which was one of Romeel's career stops, is less well known. In fact, South Africa is home to some of the best observing facilities in the Southern Hemisphere as well as having a government that financially supports astrophysics generously. Thus, South Africa is an astrophysics hotspot, also.

Results published already in Holbrook (2021), included the surprising commonality of impostor syndrome and the connection of unemployment to female scientists and not males. However, those results were connected to less than 10 interviews and further interviews included men that had been unemployed, too: but the incidence of impostor syndrome remains high.

Several foci have emerged from the interviews (thus grounded in the interviews), these foci were not part of the original project proposal, but have turned out to be significant. COVID-19 has led to stress around productivity, childcare, mental health and the cultural changed necessary for moving to predominately virtual interaction mode with physical isolation. The scientists often brought up how global pandemic is changing their lives without prompting. The scientists expressed many emotions attached to their career decision making and their professional experiences including resentment, indignation, hopelessness, apathy and depression. The astrophysics community has been drawing awareness to sexual harassment and bullying and are actively creating policies and protocols to combat these (Ball, 2021; Clancy et al., 2017; Gordon, 2016; Richey et al., 2020); however, another form of bullying was mentioned by over 10% of those interviewed: Stalking. Discussion of these new foci will be presented in future articles.

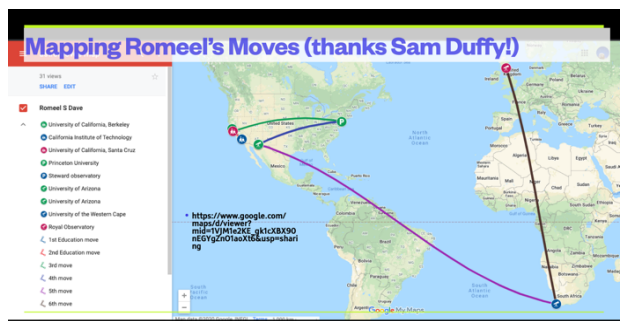

**Figure 3: The Career Moves of Professor Romeel Davé. The map was generated using Google™ tools and created by high school research assistant Sam Duffy. The moves are mapped in order. The list of institutions is on the left, with the first postdoc being at Princeton University.**

**Conclusion**

ASTROMOVES is a modern study of astrophysicists and other scientists in adjacent fields that expands cultural astronomy studies of people and their connections to the sky. The methodology is largely qualitative with interviews being the primary data source. The 35 interviews collected provide details of intersectional identities, career navigation and career decision-making. ASTROMOVES analysis has shown that factors considered during career decision-making change in weight and importance over time, such as career decisions connected to family becoming more important over time as in the case of Palolo, which is presented. Thus far, ASTROMOVES findings do not contradict the findings of previous surveys, but at the same time adds that time dimension which calls into question the previous survey results because of the changing importance of factors over time. Moving forward, analysis will include the foci that emerged from the interviews of the impacts of the COVID-19 global pandemic, the emotions connected to decision-making and career navigation, and stalking as part of the analysis of bullying and sexual harassment. Finally, a documentary film on ASTROMOVES is planned for release in 2023.

**Acknowledgements**

This project has received funding from the *European Union's Horizon 2020 research and innovation programme under the European Union's Horizon 2020 research and innovation programme* under grant agreement No786281. With support from the Institute for Advanced Studies in the Humanities (IASH) and the Department of Science, Technology and Innovation Studies (STIS), at the University of Edinburgh.

**References**

American Astronomical Society. 2021. *Current Job Ads | AAS Job Register*. https://jobregister.aas.org/

Anonymous. 2020. *AstroBetter | Rumor Mill Faculty-Staff 2019-2020*. AstroBetter. http://www.astrobetter.com/wiki/Rumor+Mill+Faculty-Staff+2019-2020

Ball, P. 2021. *Bullying and harassment are rife in astronomy, poll suggests* [News]. In: Nature. https://www.nature.com/articles/d41586-021-02024-5

C. Megan Urry, L. D., Lisa E. Sherbert, & Shireen Gonzaga (Ed.). 1993. Women at work: A meeting on the status of women in atronomy. In: *Women in Astronomy* (Vol. 1–Book, Section). Baltimore; The Space Telescope Institute.

Clancy, K. B. H., Lee, K. M. N., Rodgers, E. M., & Richey, C. 2017. Double jeopardy in astronomy and planetary science: Women of color face greater risks of gendered and racial harassment. In: *Journal of Geophysical Research: Planets*, *122*(7), 1610–1623. https://doi.org/10.1002/2017JE005256

Crenshaw, K. 1989. Demarginalizing the Intersection of Race and Sex: A Black Feminist Critique of Antidiscrimination Doctrine, Feminist Theory and Antiracist Politics. In: *University of Chicago Legal Forum*, *1989*(1), 139–167.

Crenshaw, K. (1991). Mapping the Margins: Intersectionality, Identity Politics, and Violence against Women of Color. *Stanford Law Review*, *43*(6), 1241–1299. https://doi.org/10.2307/1229039

Fohlmeister, J., & Helling, C. 2012. Career situation of female astronomers in Germany. In: *Astronomische Nachrichten*, *333*(3), 280–286. https://doi.org/10.1002/asna.201211656

Fohlmeister, J., & Helling, C. 2014. Careers in astronomy in Germany and the UK. In: *Astronomy & Geophysics*, *55*(2), 2.31-2.37. https://doi.org/10.1093/astrogeo/atu080

Gordon, O. 2016. *For Female Astronomers, Sexual Harassment Is a Constant Nightmare*. Broadly. https://broadly.vice.com/en_us/article/youre-targeted-sexually-how-female-astronomers-are-being-hounded-out-of-work

Holbrook, J. 2021. ASTROMOVES: Astrophysics, Diversity, Mobility. *ArXiv E-Prints*, arXiv:2101.10826.

Hughes, A. M. 2014. The 2013 CSWA Demographics Survey: Portrait of a Generation of Women in Astronomy. In: *STATUS: A Report on Women in Astronomy*, *January 2014*, 1–9.

IAU. 2021. *IAU Member Statistics (August 2021)*. https://www.iau.org/public/themes/member_statistics/

Ivie, R., & White, S. 2015. Is There a Land of Equality for Physicists? Results from the Global Survey of Physicists. In: *La Physique Au Canada*, *71*(2), 69–73.

Ivie, R., White, S., Garrett, A., & Anderson, G. 2013. *Women among Physics & Astronomy Faculty | American Institute of Physics*. https://www.aip.org/statistics/reports/women-among-physics-astronomy-faculty

Jacobson-Galan, W. 2021. So You Want to be a Professor of Astronomy? *Astrobites*. https://astrobites.org/2021/07/16/astronomy-professors/

Mao, Y.-Y., & Blaes, O. 1998. *Astronomy and Astrophysics Outlist*. https://astro-outlist.github.io/90s/

Ong, M. (Mia). 2001. Playing with In/Visibility: How Minority Women Gain Power from the Margins of Science Culture. In: *Women in Higher Education*, *10*(11), 42–44.

Richey, C. R., Lee, K. M. N., Rodgers, E., & Clancy, K. B. H. 2020. Gender and sexual minorities in astronomy and planetary science face increased risks of harassment and assault. In: *Bulletin of the AAS*, *51*(4). https://baas.aas.org/pub/2019i0206/release/1

Rolin, K., & Vainio, J. 2011. Gender in Academia in Finland, In: *Science & Technology Studies*, *24*(1), 26–46. https://doi.org/10.23987/sts.55268

Sean McWhinnie. 2017. *The Demographics and Research Interests of the UK Astronomy and Geophysics Communities 2016*. The Royal Astronomical Society. https://ras.ac.uk/ras-policy/community-demographics/demographic-survey-2017